\documentclass[conference]{IEEEtran}

\usepackage[normalem]{ulem}
\usepackage[hyphens]{url}
\usepackage[sort,nocompress]{cite}
\usepackage{longtable}
\usepackage{ragged2e}
\usepackage{booktabs}
\usepackage{multirow}
\usepackage{enumitem}
\usepackage[export]{adjustbox}
\usepackage{amsmath,amssymb,amsfonts}

\newcommand{\ignore}[1]{}
\usepackage[pass]{geometry}
\usepackage{fancyhdr}
\usepackage[normalem]{ulem}
\usepackage[hyphens]{url}
\usepackage{color}
\usepackage{soul}

\usepackage{fancyhdr}
\usepackage[normalem]{ulem}
\usepackage[hyphens]{url}
\usepackage[sort,nocompress]{cite}
\usepackage[final]{microtype}
\usepackage{flushend}
\usepackage{cite}
\usepackage{amsmath,amssymb,amsfonts}
\usepackage{algorithmic}
\usepackage{graphicx}
\usepackage{textcomp}
\usepackage{xcolor}
\usepackage{fancyhdr}
\usepackage{tikz}
\usepackage{siunitx}
\usepackage{gensymb}
\usepackage{amssymb}%

\usepackage{pifont}

\newcommand*\cir[1]{\tikz[baseline=(char.base)]{
            \node[shape=circle,draw,inner sep=1pt, fill=white, text=black] (char) {#1};}}

\newcommand{\xmark}{\ding{55}}

\usepackage[hyphens]{url}
\usepackage{listings}
\usepackage{subcaption}

\usepackage{hyperref}

\definecolor{dkgreen}{rgb}{0,0.6,0}
\definecolor{gray}{rgb}{0.5,0.5,0.5}
\definecolor{mauve}{rgb}{0.58,0,0.82}
\newcommand{\hpcasubmissionnumber}{1075}

\title{\vspace{-0em}DeACT: Architecture-Aware Virtual Memory Support for Fabric Attached Memory Systems\vspace{-0em}}
\author{}

\author{
\IEEEauthorblockN{Vamsee Reddy Kommareddy}
\IEEEauthorblockA{\textit{University of Central Florida}\\
Orlando, Florida, USA\\
vamseereddy8@knights.ucf.edu}
\and
\IEEEauthorblockN{Clayton Hughes}
\IEEEauthorblockA{\textit{Sandia National Laboratories}\\
Albuquerque, New Mexico, USA \\
chughes@sandia.gov}\\
\IEEEauthorblockN{Amro Awad}
\IEEEauthorblockA{\textit{North Carolina State University}\\
Raleigh, North Carolina, USA\\
ajawad@ncsu.edu}
\and
\IEEEauthorblockN{Simon David Hammond}
\IEEEauthorblockA{\textit{Sandia National Laboratories}\\
Albuquerque, New Mexico, USA \\
sdhammo@sandia.gov}
}

\begin{document}
\maketitle
\pagestyle{plain}

\begin{abstract}

The exponential growth of data has driven technology providers to develop new protocols, such as cache coherent interconnects and memory semantic fabrics, to help users and facilities leverage advances in memory technologies to satisfy these growing memory and storage demands. Using these new protocols, fabric-attached memories (FAM) can be directly attached to a system interconnect and be easily integrated with a variety of processing elements (PEs). Moreover, systems that support FAM can be smoothly upgraded and allow multiple PEs to share the FAM memory pools using well-defined protocols. The sharing of FAM between PEs allows efficient data sharing, improves memory utilization, reduces cost by allowing flexible integration of different PEs and memory modules from several vendors, and makes it easier to upgrade the system. 

One promising use-case for FAMs is in High-Performance Compute (HPC) systems, where the underutilization of memory is a major challenge. However, adopting FAMs in HPC systems brings new challenges. In addition to cost, flexibility, and efficiency, one particular problem that requires rethinking is virtual memory support for security and performance. To address these challenges, this paper presents decoupled access control and address translation (DeACT), a novel virtual memory implementation that supports HPC systems equipped with FAM. Compared to the state-of-the-art two-level translation approach, DeACT achieves speedup of up to 4.59x (1.8x on average) without compromising security.

\end{abstract}
\vspace{-1.0em}
\section{Introduction}
\label{sec:intro}

With the ever increasing demand for larger memory capacities, many high-performance computing (HPC) systems nowadays have their nodes equipped with hundreds of gigabytes of memories. For instance, Oak Ridge National Lab's Summit supercomputer has 512GB of DRAM and 96GB of HBM2 per compute node. The driver for increased memory capacity per compute node is the increasing memory needs for current applications and emerging workloads. Most HPC systems typically run many different applications from a variety of domains, each of which will have its own unique resource requirements; some applications may use the whole memory in the node while others may only use a few gigabytes. Nevertheless, most current HPC schedulers allocate resources at the node granularity and applications with extremely low memory demands can end up reserving nodes with large memories. Unfortunately, the current approach is to choose the size of memory per node based on the maximum footprint (per node) of the applications of interest, which can lead to significant under-utilization of the memory system. Moreover, applications that are not able to fit their memory needs into one node incur additional communication overhead because their computation must be split across nodes. Ideally, compute nodes should have direct access to memories that meet their demands without the need to incur expensive software stack overhead due to message passing libraries.

Recent standards, such as Gen-Z\cite{genz} and Compute Express Lanes (CXL)\cite{sharma2019compute}, define protocols and interface requirements for accessing memory modules attached to the fast system interconnect. Memory modules that implement memory-semantic protocols and can be readily integrated with the system fabric are typically referred to as Fabric-Attached Memories (FAMs). Protocols defining how to integrate FAMs are being developed through a consortium of major vendors, such as Intel, HPE, AMD, IBM, Lenovo, and VMWare\cite{sharma2019compute,genz,ccix}. FAMs promise a new HPC architecture where compute nodes can potentially access shared physical memory pools through fast interconnects. In particular, there has been recent industrial interest in architectures where memory modules can be disaggregated from compute nodes, and hence allows node to scale up its memory allocation to the requirements of the workloads run on the node. Such architectures are typically referred to as \textit{memory-centric architectures}. Memory-centric architectures leverage memory semantic protocols to communicate with FAM pools over high-speed interconnects. Memory-centric architectures promise efficiency, flexibility, and reduced costs. Examples of architectures that resemble memory-centric systems include Facebook's Disaggregated Rack\cite{facebookFabric}, HPE Labs' \textit{The Machine}\cite{machine}, and Intel's Rack Scale Architecture\cite{kyathsandra2013intel}. 

Since memory-centric architectures leverage FAMs as physically shared memory pools, multiple compute nodes, potentially running applications from different users, can access pages in the same FAM memory modules. This access model is different from conventional HPC architectures where each compute node has its own memory modules and applications' memory accesses are limited to its own nodes. Therefore, a new question arises, who is responsible for access control of FAMs? Without strict access control mechanisms, malicious operating systems (OSes), applications, and processing elements (e.g., accelerators, SoCs, etc.) can potentially compromise the entire system by accessing the data of other users in the shared FAMs. Note that in this system architecture, there could be compute nodes containing processing elements (PEs) from different vendors. Even if not malicious, these PEs could contain bugs in their internal virtual memory implementation, which compromise the whole system. Obviously, with such a wide attack surface, accesses to shared FAM modules need to be vetted externally, at the system-level, and not rely solely on access control within the PEs. Pages in shared FAM pools can be managed in two different ways. The first approach is through transparently allocating FAM pages to nodes on-demand, i.e., each compute node has the illusion that it has a contiguous large physical space \cite{lim2009disaggregated}. Such approach is similar in spirit on how hypervisors give virtual machines (VMs) the illusion that each VM has a contiguous guest physical address, which eventually gets translated into the real system physical address through hypervisor. In our case, a memory broker node is dedicated to set up such translations for each node at system level. The second approach is to expose each node to the real physical addresses (FAM addresses) and modify the kernel running on each node to communicate with external memory broker to allocate FAM pages \cite{lim2009disaggregated,shan2018legoos}.

Transparent management of FAMs' pages eliminates the need to modify the kernel and, most importantly, allows system-level vetting of accesses to FAMs through a second level of memory translation, from the node guest address to the FAM physical address. However, while this is similar to two-level translation in virtualized environments, the flexibility, transparency and security come at the cost of significant performance overheads due to the additional level of translation. In conventional x86 systems, each memory access can require up to four memory accesses for translation, however, when a second level is added, the number of memory accesses can be up to 24\cite{bhargava2008accelerating}. We observe that significant performance overheads can be incurred when naively implementing state-of-the-art implementations of two-level translation, using system translation unit (STU) as shown in Figure \ref{fig:expo-ind}(b) for memory-centric systems. STU is similar in spirit to \textit{Gen-Z memory management unit (ZMMU)} \cite{ZMMU}. Hence, in this paper, we focus on optimizing the implementation of virtual memory support for memory-centric systems.  

To minimize the performance overheads of transparent access control and management support for shared FAM pools, we propose \textit{decoupled access control and address translation (DeACT)}. DeACT leverages the architecture layout of memory-centric architectures and the ability to decouple access control from translation. Specifically, DeACT allows \textit{unverified} caching of translations in the small local memories within compute nodes, but enforces access control at the system level. By decoupling access control and translation and leveraging part of the local memories in compute nodes as unverified caches, DeACT exploits high spatial locality of access control metadata for each node. This architecture-aware decoupling and unverified caching brings significant performance improvements and reduces the number of translation requests significantly, while strictly enforcing access control.

Previous proposals discussed virtual memory support for remote memory architectures \cite{shan2018legoos,aguilera2017remote,aguilera2018remote,lim2009disaggregated}. However, these techniques significantly modify the OS \cite{shan2018legoos}, propose methods which are non transparent to the applications requiring programmer intervention \cite{aguilera2018remote}, limit the use of remote memory merely as a swap space \cite{lim2009disaggregated} or invalidate virtual memory paging for FAM architectures \cite{aguilera2017remote}. None of these approaches consider the performance impact of address translations in FAM. Also, these methods do not consider security aspects. In contrast, we provide quantitative analysis of address translation overheads for FAM systems and propose DeACT, a virtual memory support mechanism that leverages the architectural layout of FAM systems to use in-memory caching, that greatly exceeds the capacity of on-chip caches, without modifying OS. 

To evaluate our scheme, we use the Structural Simulation Toolkit (SST)\cite{rodrigues2011structural}, a publicly available architectural simulator. We leverage the currently existing disaggregated memory model and memory manager, Opal\cite{kommareddy2018opal}, to model our system. We evaluate DeACT for applications using multiple memory-intensive benchmark suites. DeACT achieves speedups of up to 4.59x (1.8x on average), compared to the state-of-the-art two-level translation approach, without compromising security.

The organization of the paper is as follows. First, we discuss the threat model, memory management in FAM systems and motivation in Section \ref{sec:background}. Section \ref{sec:design} explains how access control is decoupled from address translations and how unverified caching operate in local memory. Methodology and applications are discussed in Section \ref{sec:method}. Results and sensitivity analysis are discussed in Section \ref{sec:results}. Section \ref{sec:disc} discusses feasibility of huge pages and various other factors in FAM systems. Finally, we walk through the related work in Section \ref{sec:related} and conclude in Section \ref{sec:conclusion}.
\vspace{-3mm}
\section{Background}
\label{sec:background}
In this section, we discuss our assumed threat model, memory management concepts in FAM and relevant systems and FAM.

\subsection{Threat Model}
\label{sec:threat}
\vspace{-2mm}
In our threat model, we assume that compute nodes themselves can have bugs which can be exploited by malicious applications or OSes. Such threats are common and evidenced by the recent vulnerabilities in Intel's processors (e.g., Meltdown \cite{lipp2018meltdown} and Spectre\cite{kocher2019spectre}). Our threat model assumes that a malicious application or OS runs on a specific node which tries to illegitimately access memory pages of other nodes and users in the shared FAMs. Note that we assume a compute node, at any point of time, is owned by a single user (i.e., a user allocated the node to run an application). By exploiting a bug in virtual memory implementation within a compute node or a vulnerability in OS, the attackers can directly map its own virtual space to any physical page in the shared FAM and hence be able to access it freely. Therefore, to minimize the attack surface, an additional level of access control needs to vet accesses that come from compute nodes to ensure that they are for pages belonging to the node. Thus, any pages in FAM that are considered exclusive to a node, must be protected from any access by other nodes as long as such page is allocated. Attacks such as timing side-channel, covert-channel and physical attacks are beyond the scope of this paper and can be addressed with many available solutions based on the system nature. Similar to most HPC systems, we disallow co-locating resource allocations on the same node, which minimizes the risk of information leakage (within a node). 
In summary, we mainly focus on enforcing access control on shared FAMs, to limit the impact of vulnerabilities within compute nodes on other compute nodes' data. Such protection is analogous to security guarantees provided by virtual memory for applications running on a native system, but at node level. Our threat model trusts the memory and fabric, i.e., memory provides nodes with the requested data and does not try to give them data of other locations. Similarly, the fabric will not change the address in a request after it has been vetted by access control.


\subsection{Hierarchical Page Tables}
\vspace{-2mm}
Hierarchical (multi-tier) page tables are commonly used in modern servers due to their performance, dynamic growth, and scalability. In such settings, a virtual address is provided as an input to the translation process, then the offsets (derived from virtual address) are used to index each level to obtain the address of the next level. Finally, the last level, typically called Page Table Entry (PTE) level, has the actual translation entry, i.e., the corresponding physical address and the access permissions. However, to reduce translation overheads, hardware-support for memory management, typically implemented as the Memory Management Unit (MMU) and maintained by the OS, is provided. The MMU is responsible of caching the translations, i.e., PTEs, in TLBs. Moreover, MMU can also cache the contents of different levels of the page table in what is called as page table walking (PTW) caches \cite{bhargava2008accelerating}. Finally, MMU is responsible for walking the page table in case of TLB miss to complete the translation process. 

\vspace{-0mm}
\begin{figure}
    \centering
    \includegraphics[width=1\columnwidth]{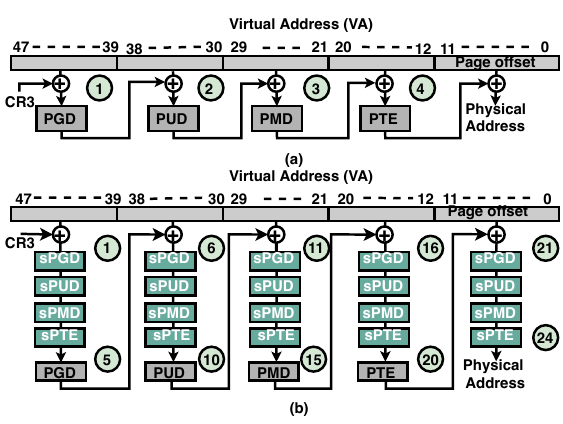}
    \caption{Page table walking (a) x86 and (b) virtualized systems.}
    \label{fig:ptw-nested}
\end{figure}

As shown in Figure \ref{fig:ptw-nested}(a), for every TLB miss, the page table is walked. In x86-64 systems, a 4-level page is typically used, and the levels are called PGD, PUD, PMD and PTE, respectively. Therefore, each memory access needs additional four memory requests to walk the page table. The root of the page table of the process currently executing on the core is loaded in Control Register 3 (CR3) in the core. In each virtual address, the bits beyond the page offset (typically 12 bits) are divided into multiple sections (9 bits each) where each section is used to index a specific level in the page table. Finally, the last-level page (PTE) has the actual page mapping. Hence, for every TLB miss, the entire page table is walked which will incur an additional four memory accesses.

In virtualized systems, hypervisors like Xen\cite{xi2011rt} are responsible for maintaining multiple guest systems and managing their memory.
For such systems, nested paging is one approach in which two page tables are maintained. One to convert virtual address to guest address and the other is a nested page table to converts guest address to system physical address. Each level of the guest page table has to walk the nested page table which requires four more memory accesses per guest page table level, Figure \ref{fig:ptw-nested}(b). Hence, 24 memory accesses are required to fetch the translation. This leads to huge overhead and \cite{bhargava2008accelerating} proposed PTW caches, nested TLBs and nested PTW caches to reduce the number of memory accesses to translate an address. PTW cache unit caches the intermediate level address translations and helps in reducing the average number of PTW steps.

\vspace{-1mm}
\subsection{Fabric Attached Memory (FAM)}
\label{ssec:mmdms}
\vspace{-2mm}
Recently, there has been a push towards disaggregating resources (such as memory modules) \cite{shan2018legoos,aguilera2017remote,aguilera2018remote,lim2009disaggregated, lim2012system,kommareddy2019page} and connecting them through system fabric \cite{gu2017efficient,gen2019gen,ccix,birrittella2015intel}. Fabric-attached memories (FAMs) are memory modules that are attached to system fabric which can be accessed by multiple nodes through conventional load/store operations. FAMs allow disaggregating memory pools and make them accessible to compute nodes through a fabric using well-defined memory access protocols, such as Gen-Z\cite{genz}, CXL\cite{sharma2019compute}. Managing disaggregated FAM uniquely combines the challenges of managing shared FAM and multi-level memory (local memory and remote memory). On one hand, nodes should be prohibited from illegitimately accessing data belong to other nodes. On the other hand, OSes running on nodes are agnostic to the actual status of the FAM and thus would instead manage a hypothetical large node physical memory which translates into FAM. However, to enforce access control and permissions, translating nodes' physical address into FAM address should occur at the system-level (off-the node), not at the node-level. By doing so, any malicious OS or node, will be limited to accessing the data allocated to the node instead of the whole global memory.

\begin{figure}
    \centering
    \includegraphics[width=1\columnwidth]{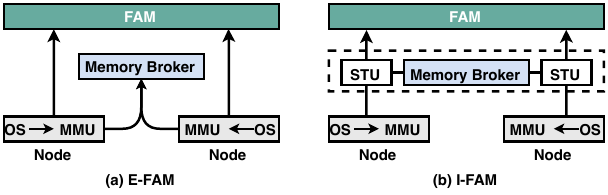}
    \caption{Two ways of managing memory in FAM systems.}
    \label{fig:expo-ind}
\end{figure}

Adding a translation layer at system-level allows running unmodified OSes on nodes and enforcing access control, but incurs significant performance overheads. Lim et al. \cite{lim2009disaggregated} explored two-stage address translation system for disaggregated memories. For the rest of the paper, we call such a scheme \textit{Indirect FAM (I-FAM)}, since FAM is accessed indirectly. In I-FAM, a simple STU can be implemented in a router connected directly to the node or in the memory blade, Figure \ref{fig:expo-ind}(b). STU is responsible of caching system-level translations, i.e., node address to FAM address, and access permissions. Moreover, STU is capable of sending address translation service requests (similar to PCIe's \cite{ATS}) or request physical pages from the system-level memory broker (in case of unmapped addresses). Note that STU is similar in spirit to the ZMMU \cite{ZMMU}. Exposing the FAM directly to nodes' OSes would significantly improve the performance by removing the additional translation layer, Figure \ref{fig:expo-ind}(a). We refer to such a scheme as \textit{Exposed FAM (E-FAM)}. In E-FAM, OSes need to be patched to communicate with global memory manager node (e.g., through MPI interface) to coordinate memory management with other nodes. While  E-FAM provides translation overheads as low as native systems, it requires modifying OS \cite{shan2018legoos} and enormously enlarges the attack surface; any malicious node/OS can map its address space into any location in global memory and hence leaks data from other nodes.

\vspace{-2mm}
\subsection{Motivation}
\label{sec:motivation}
\vspace{1mm}
\begin{figure}[htbp!]
    \centering
    \includegraphics[width=1\columnwidth]{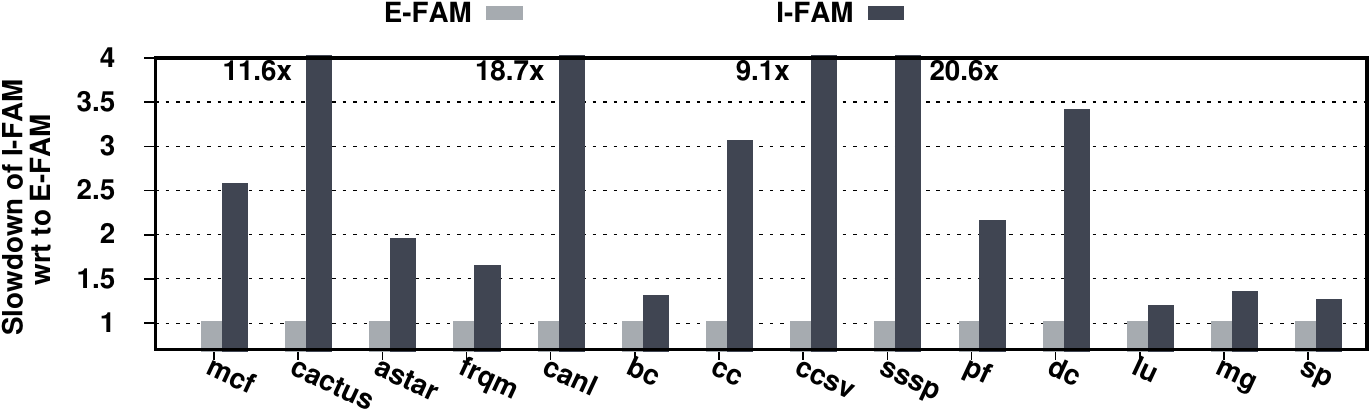}
    \caption{Normalized performance with respect to E-FAM.}
    \label{fig:vm-motivation-ipc}
    \vspace{-4mm}
\end{figure}
\vspace{-3mm}
\begin{figure}[htbp!]
    \centering
    \includegraphics[width=1\columnwidth]{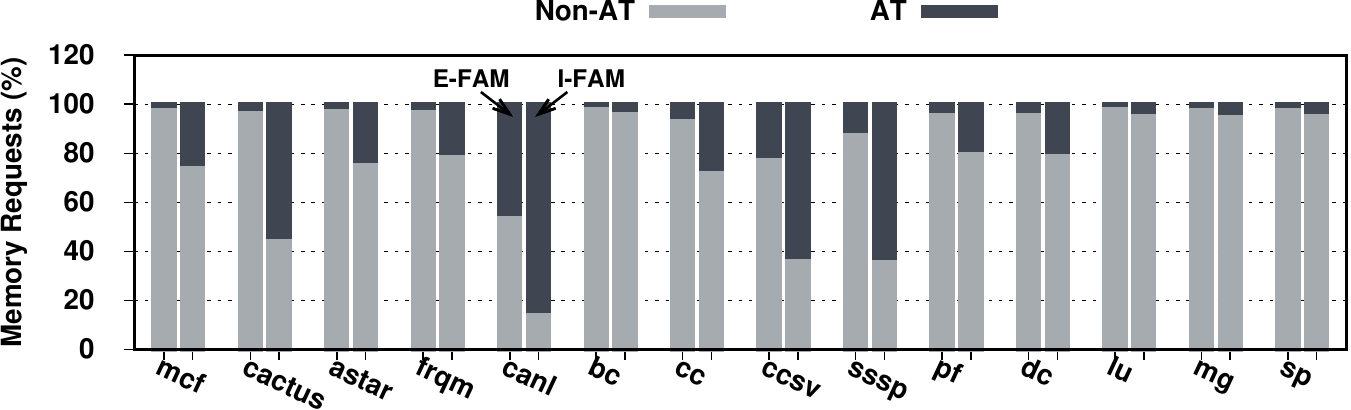}
    \caption{Breakdown of percentage of address translation (AT) and non address translation (Non-AT) requests observed at FAM in I-FAM and E-FAM systems.}
    \label{fig:vm-motivation-mem}
    
\end{figure}
Figure \ref{fig:vm-motivation-ipc} shows slowdown in I-FAM compared to insecure E-FAM wherein no indirection is needed. We observe a performance degradation of 20.6x for \textit{sssp} benchmark\footnote{Details about the methodology and the benchmarks are in Section \ref{sec:method}.}. The slowdown is attributed to the increased address translation requests observed at FAM due to indirection at system-level, Figure \ref{fig:vm-motivation-mem}. For instance, the percentage of address translation requests for \textit{canl} benchmark is 44.36\% in E-FAM. However, this increases to 84.13\% in I-FAM. Also, we note that benchmarks which are not sensitive to address translations become highly sensitive to address translations in I-FAM. Address translation requests increase from 1.81\% to 53.69\% for \textit{cactus} benchmark. Clearly, I-FAM brings in significant performance overheads. The goal of DeACT is to design a FAM architecture with better performance without sacrificing security and with minimum or no changes to OS. A comparison of baseline FAMs and our proposed DeACT FAM approach is shown in Table \ref{table:vm-arch}.

\vspace{-3mm}
\begin{table}[htbp!]
\small
   \caption{FAM Architectures Comparison.}
  \centering
 
  \begin{tabular}{|p{1.6cm}|p{1.6cm}|p{2.6cm}|p{1.1cm}|}
    \hline
    \textbf{Architecture} &\textbf{Performance} &\textbf{Avoid OS Changes}
    &\textbf{Security}\\
    \hline
    \hline
      E-FAM & \checkmark & \xmark & \xmark  \\
     \hline
      I-FAM & \xmark & \checkmark & \checkmark  \\
    \hline
      DeACT & \checkmark & \checkmark & \checkmark\\
    \hline
  \end{tabular}
     \label{table:vm-arch}
     \vspace{-3mm}
\end{table}

\section{Design}
\label{sec:design}
In this section, we describe our proposed approach DeACT to provide virtual memory support in FAM systems.

\vspace{-1mm}
\subsection{DeACT Overview}
\label{sec:vm-deact-overview}
\vspace{-2mm}
When designing support for virtual memory, we aim at abstracting away the details of the global memory from nodes' OSes, however, while enforcing isolation and minimizing translation overheads. To do so, we adopt a two-layer approach where each node's OS manages an imaginary flat node physical memory. The node physical memory range can be thought of as a range of two different NUMA zones, one zone (low addresses) corresponds to the local DRAM and the other zone (high addresses) corresponds to the FAM. With such a design, each node's OS manages its memory allocations oblivious to the actual status of FAM. While such an approach abstracts away the complexity of managing a shared resource (memory), it adds significant performance overheads due to two levels of indirection.
Therefore, we need novel mechanisms to improve the performance of such design. 

One major observation we make is that access control can be decoupled from the translation process. In particular, the translation from node address to FAM address can be sped up significantly by caching the translations at node-level memory. Later, if the translation of a specific node address exists locally in the node, the global memory request is forwarded to the FAM, with the obtained/cached FAM addresses. In other words, the node can provide the final FAM addresses it needs to access. As the reader can expect, the access control is offloaded to the off-the node components, e.g., STU units. Since the access permissions need to be checked for the specific FAM address provided by the node, we dedicate specific parts in the FAM to store FAM access control metadata (ACM). Such parts are known for STUs and the addresses of the ACM of any FAM page can be derived merely from the FAM address. For instance, assume we want to keep a 16-bit ACM for each 4KB page and assuming the metadata starts at address $MTAdd$ in FAM. To read ACM of FAM address \textit{X}, we read the 64-byte block at address $MTAdd+\frac{X}{4096 \times 32}$. For simplicity, the metadata of each 4KB page is node ID of the node that owns that page and read, write and execute permissions. Read, write and execute fields consumes two bits and the rest of the bits (14) are allocated for the node ID. FAM pages could also get shared between the nodes. Thus, we use all the node ID bits of the page metadata set to 1 to indicate a shared page. Hence, we can have up to 16383 nodes supported in the system.

Since pages can be shared by a subset of nodes, just indicating a page is shared is insufficient. Therefore, we use a bitmap-like scheme to indicate which nodes are allowed to access a specific page. However, since having a bitmap for each 4KB can introduce significant overheads, we limit shared pages to 1GB physical pages. For each 1GB physical page in global memory, we have a corresponding 64K bits bitmap (8KB) in the metadata region. Since such overhead is negligible (less than 0.0001\%), and to enable easier indexing of metadata, we dedicate a bitmap for each 1GB physical region regardless of being used as a shared page or not. Therefore, when ACM is accessed, if the node ID bits of the metadata indicates a shared page, we immediately fetch the corresponding parts of the bitmap to check if the node has access permissions. In contrast, if the node ID bits does not indicate a shared page, we simply compare its value (owner node ID) with the ID of the requesting node, to verify the legitimacy of the access. Note that when a shared page is allocated (or becomes shared), all of its node ID bits in the metadata fields correspond to its 4KB chunks (sub-pages) are set to shared, i.e., {\tt 0xfffd}. When the page is shared the last two bits of the metadata field indicate read, write and execute permissions assigned to the node. This enable enforcing mixed access permissions for nodes sharing a page. For instance specific subset of nodes are allowed to read and write to the shared page and the rest of them only read the shared page.

\begin{figure}
    \centering
    \includegraphics[width=0.7\columnwidth]{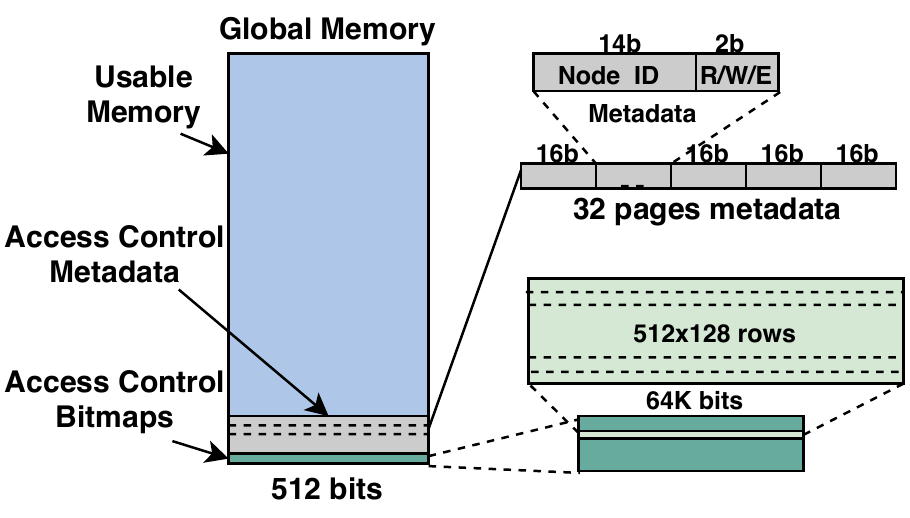}
    \caption{Page access control metadata and bitmaps in FAM.}
    \label{fig:metadata}
    \vspace{-8mm}
\end{figure}

One obvious optimization to reduce the overheads of obtaining such ACM is to cache them. However, since such metadata must be enforced by FAM managers, not the node, such metadata should be only cached outside the nodes and inaccessible by the nodes or their own OSes. Therefore, we opt for caching such metadata in STUs. Such STUs can potentially have a small lookup table, similar to TLBs. As mentioned earlier, such STUs can be added to the global memory blade or simply at the first router/switch that connects a node to the system fabric. It is also important to note that such metadata has very high spatial locality, a single 64B block covers 32 4KB pages, i.e., 128KB region for a 16-bit ACM. Therefore, even a very small TLB-like cache can save significant number of reads to access control metadata. While beyond the scope of this paper, in encrypted memories, if each node uses a unique memory encryption key, we could allow read requests without checking access control; writes can tamper with data but reads are useless if the node has different unique key, and thus no need for enforcing access control for reads.

As we now understand how our decoupled access control works, we will discuss how we accelerate the translation process. To speed up the translation process, we (a) propose node-level unverified caching of system-level translations: We notice that a very small portion of local memory can be used to cache system-level translations, which will be later sent for verification at system-level. (b) efficiently cache ACM in STU. 

\subsection {System Overview}
\label{ssec:deact}
\label{ssec:vm-2-step}
\label{ssec:vm-fam-ptw}
\begin{figure}
    \centering
    \includegraphics[width=1\columnwidth]{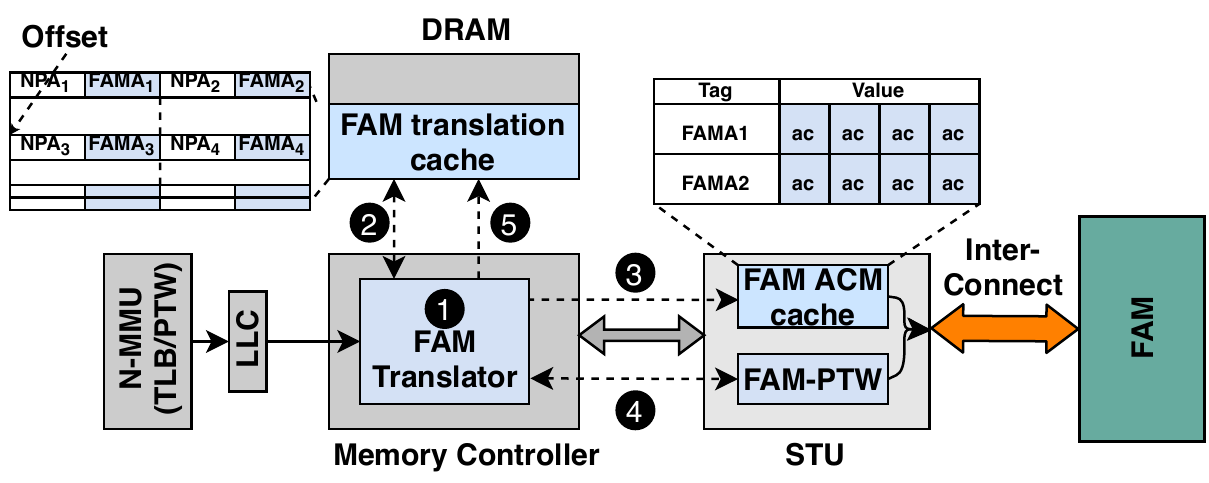}
    \caption{DeACT FAM schematic.}
    \label{fig:vm-schematic}
    \vspace{-8mm}
\end{figure}

\vspace{-1mm}
Figure \ref{fig:vm-schematic} shows a schematic overview of our proposed design. Decoupling the metadata from page mapping qualifies the system-level translations to be cached in the local memory. Hence we maintain a FAM translation cache in the DRAM. We add a \textit{FAM translator} \cir{1} in the memory controller to map node addresses to FAM addresses by accessing the FAM translation cache \cir{2}. Although node addresses are mapped to FAM addresses by the FAM translator unit it is still a partial translation since accesses have to be verified. To complete the mapping, the FAM accesses are verified by the STU \cir{3}. Hence unlike I-FAM, it requires two steps to translate a node address and verify the access.

For a translation miss the FAM page table has to be walked. In our design we let the walking to be done by the STU since we observe that the overhead of including FAM PTW inside the node is costlier than the benefits. Firstly, due to security reasons, we aim at clean separation of address translations and ACM, if the fabric translations are cached inside the node. Hence if the intermediate translations are also cached within the node, the ACM for intermediate page tables should also be decoupled. However, the two step process required to complete the mapping, delays address translations significantly, considering four memory accesses during PTW. That is at every intermediate level the ACM should also be fetched from the memory incurring additional memory accesses. Secondly, since FAM translation cache size in local memory is significantly higher that the STU cache size, we observe a hit rate of more that 90\% in FAM translation cache in the local memory. Hence walking the FAM page table within the node would unnecessarily increase the complexity without much benefits. Also, it would increase the complexity of the memory controller. Thus we apply DeACT only to the last level of the page table (PTE). Therefore during a FAM translation miss, FAM translator forwards the missed request to the STU, which walks the FAM page table and fetches the entry on behalf of FAM translator \cir{4}. After receiving the missed translation, FAM translator maps the pending requests and then updates FAM translation cache in the local memory\cir{5}.

\subsection {FAM Translator}
\label{ssec:vm-fam-tr}
\vspace{-1mm}
The idea of FAM translator in the local memory controller is to translate node addresses to FAM addresses without verifying memory accesses. Functionalities of FAM translator are: (a) fetching the translation from the FAM translation cache (b) matching the tag (c) handling translation hits and misses (d) handling off-the node responses and (e) updating FAM translation cache in the local memory.

\begin{figure}
    \centering
    \includegraphics[width=1\columnwidth]{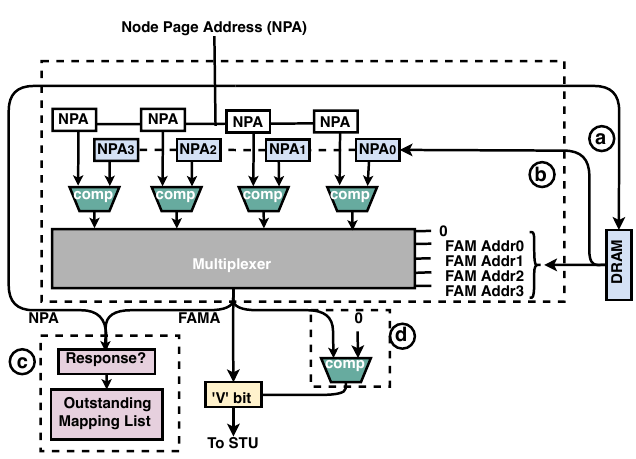}
    \caption{FAM translator.}
    \label{fig:vm-fam-tr}
    \vspace{-6.5mm}
\end{figure}

\noindent \textbf{Accessing DRAM for Translation:}
To fetch the translation from the local memory, \cir{a} in Figure \ref{fig:vm-fam-tr}, the FAM translator calculates the local memory address by adding starting address of FAM translation cache to the offset, Figure \ref{fig:vm-schematic}. Offset is dependent on the type of the FAM translation cache in the local memory. For simplicity, we use a four way associative cache. This is because memory access granularity is 64-bytes and each mapping entry requires 104 bits; 52 bits for tag (node page) and 52 bits for value (FAM page), for a page size of 4KB. A single memory access fetches four entries. Thus offset is obtained by performing a modulus operation on node page number with the number of FAM translation cache sets.

\noindent \textbf{Tag Matching:}
After fetching the translations from the local memory, FAM translator matches address tags using comparators, \cir{b} in Figure \ref{fig:vm-fam-tr}. 
We add four comparators and a multiplexer to perform tag matching concurrently. If none of the tags match, the output of the multiplexer is set to 0. This takes just one cycle to match the tag but the number of comparators required are 4x more compared to using just one comparator when four tags are matched serially in four cycles. However, these additional comparators adds up minutely to the overall hardware cost and area.

\noindent \textbf{Handling Translation Hits:}
When any of the tags, fetched from the FAM translation cache, match with the required node page address the multiplexer outputs the respective FAM address. FAM translator replaces the node address with the FAM address and forwards the request. However, before forwarding the request to STU, FAM translator identifies if the request is expecting any response back from the FAM. If so the FAM address to node address mapping is stored in \textit{outstanding mapping list}, \cir{c} in Figure \ref{fig:vm-fam-tr}. This is because, FAM responses contains data tagged with FAM addresses and the node only deals with node addresses. \textit{Outstanding mapping list} is used to convert FAM address to node address during FAM response. Since the number of outstanding requests are limited (128 requests) the number of entries in \textit{outstanding mapping list} are also limited. In I-FAM this list is maintained in STU. But since the FAM translations are performed within the node and STU does not understand node addresses in DeACT, this list is maintained by the node.



\noindent \textbf{Handling Translation Misses:}
A translation is identified as a miss if the output of the multiplexer is zero, \cir{d} in Figure \ref{fig:vm-fam-tr}. During a miss FAM translator forwards the missed request to STU to walk the page table. FAM PTW unit of STU retrieves the node address from the missed request and walks the page table. After the page table is walked, STU translates the node address to the FAM address and then verifies the access to forward the missed request to the FAM. Also, STU sends the the page mapping to FAM translator for updating the FAM translation cache and to register the mapping in the \textit{outstanding mapping list} if needed.

STU receives two types of requests from a node, mapped and not mapped requests. Mapped requests are those whose node address is translated to FAM address by FAM translator. For such requests STU verifies FAM access permissions. On the other hand STU walks the page table for not mapped requests using the node address from the request address field. To make STU distinguish between the two types of requests we add a verification ('V') flag to the request packet. This flag is set by FAM translator unit if the mapping is successful and is reset for a missed translation. Using 'V' flag STU either forwards the request to the verification unit or to the PTW unit.

\noindent \textbf{Handling off-the node memory responses}
FAM translator segregates responses into two types (a) memory response and (b) mapping response. Memory responses are forwarded to the last level cache by fetching the node address from the \textit{outstanding mapping list}. Mapping responses are received from the STU PTW unit. FAM translator updates the FAM translation cache during mapping response.

\noindent \textbf{Updating FAM Translation Cache}
Since the granularity of memory access is 64 bytes each access to FAM translator operates on four FAM mapping. To update the FAM translation cache, FAM translator has to write to one of the four mappings fetched. Hence during a translation cache update, FAM translator reads 64 bytes from the local memory, updates one of the entries and writes back 64 bytes. For simplicity, we randomly selected one of the four entries to replace. It is possible to implement different cache replacement policies but such policies require additional DRAM space; to store mapping status, and additional writes to the DRAM; to update mapping status for every FAM access. 



\subsection{FAM Access Verification}
\vspace{-2mm}

\begin{figure}
    \centering
    \includegraphics[width=1\columnwidth]{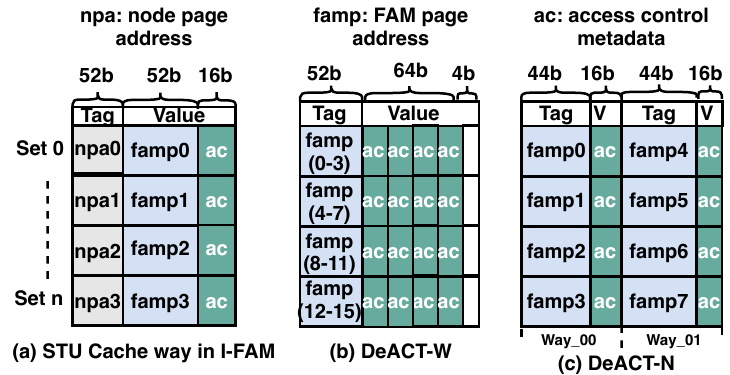}
    \caption{ACM organization in STU cache way in (a) I-FAM and (b,c) DeACT.}
    \label{fig:vm-ac-metadata}
    \vspace{-6mm}
\end{figure}

Memory accesses verification is performed by STU in DeACT. STU  verifies the memory access by a) checking if the page being accessed is assigned to the node, using node ID in the metadata and b) checking the read, write and execute permissions from the metadata. However, the metadata is not provided by the node since we maintain the page metadata in the memory off-the node to provide security, Section \ref{sec:vm-deact-overview}. Hence for STU to verify the FAM access it needs to fetch the page metadata from the memory, Figure \ref{fig:metadata}. 

Since STU is off-the node, it can be used to cache ACM. STU in I-FAM caches both FAM page mapping and ACM together, Figure \ref{fig:vm-ac-metadata}(a) (52 bits for the tag (node page address) and 52 bits for FAM page address and 16 bits for ACM). However, STU in DeACT caches only ACM. Although DeACT reduces the frequency at which the page table is walked by leveraging local memory to store FAM mappings, it introduces an additional memory access for the ACM. Hence to reduce the number of accesses to FAM for ACM, we explore organizing ACM in the available space, after decoupling the page mapping from the STU cache in DeACT.

\noindent \textbf{Way-level contiguous organization (DeACT-W)}
\label{ssec:vm-deact-w}
This is a simple organization wherein the space available in each cache way, after removing the address mapping, is used to cache ACM of contiguous pages, Figure \ref{fig:vm-ac-metadata}(b). Since ACM is 16 bits and the space available is 52 bits (FAM page address), four contiguous pages ACM is stored in one cache way. For instance ACM for pages from 0 to 3 are stored as one cache way and 4 to 7 are stored in a different cache way. Hence caching of ACM increases by four times.

\begin{figure}
    \centering
    \includegraphics[width=1\columnwidth]{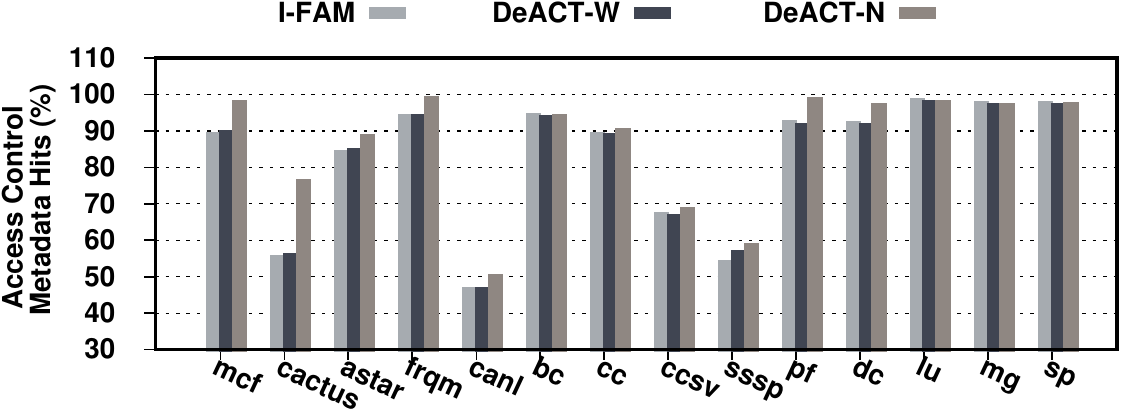}
    \caption{Access control metadata hit rate.}
    \label{fig:vm-rand-acm}
    \vspace{-8mm}
\end{figure}

\noindent \textbf{Non-contiguous organization (DeACT-N)}
\label{ssec:vm-deact-n}
With DeACT-W we observe ACM hit rate of almost 90\% for most of the benchmarks Figure \ref{fig:vm-rand-acm}. However, ACM hit rate for benchmarks like \textit{canl}, \textit{sssp} and \textit{cactus}, which are sensitive to address translations, is less than 60\%. This is because STU in DeACT-W achieves higher hit rate when spacial locality, while accessing memory, is higher. However, since FAM is shared by multiple nodes, memory allocation is random and hence has poor spacial locality while accessing memory. Thus, instead of organizing STU to cache contiguous page ACM, we organize STU to cache ACM of non-contiguous pages i.e., the free space is used to store tag and ACM pair of another page which is either contiguous or non-contiguous, Figure \ref{fig:vm-ac-metadata}(c). 

Since each tag and ACM pair needs 68 bits; 52 bits for the tag and 16 bits for the ACM, the available free space, 52 bits, is not sufficient to store an additional pair. Thus to fit ACM for two different pages within the same way of a set (within the available space) we confine the number of tag bit to 44. With 44 tag bits STU can cover 8 tera page metadata\footnote{Note that these numbers are based on the tag and data bits assumed in the STU cache of I-FAM as shown in Figure \ref{fig:vm-ac-metadata}(a)} and hence each node can access 32 petabytes of memory unlike 16384 petabytes in DeACT-W. However, 32 petabytes is also significantly higher for a node. Thus each cache way is sub divided into two sub-ways (way\_00 and way\_01 for way 0) and each sub-way has a tag and ACM. This increases the total number of ways for a set and matching the tags of sub-ways is similar to  matching the tags of different ways in a cache. Organizing ACM in STU cache in this manner doubles the caching of ACM and unlike DeACT-W, DeACT-N stores non contiguous page ACM.

Non-contiguous organization of ACM in the STU cache increases the hit rate from 90\% to almost 99\% for most of the applications. Also, hit rate for address translation sensitive benchmark, for instance \textit{cactus}, increases from less than 55\% to almost 76\%. The improvement compared to DeACT-W is due to random accesses to FAM, see Section \ref{sec:results}.

\vspace{-2mm}
\section{Methodology}
\label{sec:method}
\vspace{-2mm}
To evaluate our design we used a decoupled memory model implemented in Structural Simulation Toolkit (SST)\cite{rodrigues2011structural,kommareddy2018opal}. SST is an event-based cycle-level simulator which has been proven to be one of the most reliable simulators for large-scale systems due to the scalability and modular design of its components. SST includes multiple simulation modules for various components. To evaluate FAM architectures a FAM manager (memory broker), Opal \cite{kommareddy2018opal}, was developed in SST. We modified SST memory management unit, Samba \cite{awadsamba} and Opal \cite{kommareddy2018opal} modules to model our design. We modelled an STU component in SST to translate node addresses to fabric addresses and to verify FAM accesses. As our approach focuses on accelerating the address translations, we validate our approach by calculating the performance of the system in-terms of instructions per cycle.

\begin{table} 
    \vspace{0em}
    \centering
    \caption{System Configuration}
    \label{tab:simulationparameters}
    \scriptsize
    \begin{tabular}{|l|p{5cm}|}
    \hline
    \multicolumn{2}{|c|} {\bf Node}
    \\ \hline                     
    CPU & 4 Out-of-Order cores, 2GHz, 2 issues/cycles, 32 max. outstanding requests
    \\ \hline
    TLB & 2 levels, L1 size: 32 entries, L2 size: 256 entries
    \\ \hline
    L1 & Private, 64B blocks, 32KB, LRU                 
    \\ \hline
    L2 & Private, 64B blocks, 256KB, LRU                         
    \\ \hline
    L3 & Shared, 64B blocks, 1MB, LRU
    \\ \hline
    Local memory & DRAM, Size: 1GB                          
    \\ \hline
    \multicolumn{2}{|c|}{\textbf{STU}}
    \\ \hline
    Cache & Size: 1024 entries, associativity: 8
    \\ \hline
    \multicolumn{2}{|c|}{\textbf{Fabric Network}}
    \\ \hline
    Latency & 500ns
    \\ \hline
    \multicolumn{2}{|c|}{\textbf{Fabric Attached Memory (NVM)}}               \\ \hline
    Capacity       & 16GB                                           \\ \hline
    Latnecy & Read 60ns, Write 150ns
    \\ \hline
    Banks & 32
    \\ \hline
    Outstanding requests & 128
    \\ \hline
    \end{tabular}
    \vspace{-6mm}
\end{table}

Table ~\ref{tab:simulationparameters} shows system simulation parameters. We simulated 4 cores and each core can serve up to two instructions per cycle with a frequency of 2GHz. Each core is configured to execute a minimum of 100 million instructions of an application execution during its HPC-relevant kernels. L1, L2, and L3, caches are inclusive with sizes 32KB, 256KB and 1MB respectively. Local memory, is 1GB DRAM\cite{shan2018legoos} and FAM is 16GB NVM\footnote{In reality local memory is in GBs and global memory is in TBs or PBs. However, given slow simulation speeds, we scale down the memory sizes and among the total applications memory (average of 309MB during the simulation period), 20\% is allocated from the local memory and 80\% is allocated from the FAM. 
}. We simulated fabric network to connect to NVM memory with a network latency of 500ns, modeled after recent research and public projections for a fabric interconnects \cite{shan2018legoos,gao2016network,shrivastav2019shoal}. Two levels of TLB's, each of which are simulated with 32 and 256 entries within the node. Since STU is an external hardware per node we have restrictions over adding additional hardware. Hence to avoid significant hardware overhead we implemented STU to cache 1024 page table entries with 128 sets and associativity of 8, similar to Haswell Xeon L2 TLB design\cite{gras2018translation}. However, we also evaluate with DeACT by varying STU cache size. For optimization proposed by \cite{bhargava2008accelerating} we used 32 PTW cache entries. The proposed FAM translation cache size in DRAM is 1MB.



Since our focus is on HPC applications we evaluated benchmarks from different benchmark suits, as shown in Table \ref{tab:vm-applications}. Our selection of benchmarks from these benchmark suits are based on: (a) the benchmark should have a minimum of 5 misses per kilo instructions (MPKI) (b) should be compatible to the simulation setup (c) the performance degradation with I-FAM should be more than 15\% compared to E-FAM, since we observe application which do not get impacted much by introducing indirection degrades it performance with DeACT, explained in Section \ref{sec:results}. Considering this criteria we have evaluated 29 benchmarks and zeroed in on 14 benchmarks which are meeting the requirements. Selected benchmarks with their respective MPKI are shown in Table \ref{tab:vm-applications}. Due to the limited space we use short forms to represent applications. The short forms are next to the applications in Table \ref{tab:vm-applications}. Connected components graph analytic benchmark has 2 variants \textit{cc}; which uses Afforest sub-graph sampling algorithm \cite{sutton2018optimizing}, and \textit{ccsv}; which uses Shiloach-Vishkin algorithm \cite{shiloach1980log}.

\begin{table}
\centering
\caption{Applications}\label{tab:vm-applications}
\scriptsize
\begin{tabular}{|p{2cm}|p{3cm}|p{1cm}|} \hline
      \multicolumn{1}{|p{2.5cm}|}{\multirow{1}{*}{\textbf{Benchmark Suite}}}
    & \multicolumn{1}{|p{3cm}|}{\textbf{Application}} 
    & \multicolumn{1}{|p{1cm}|}{\textbf{MPKI}} \\ \cline{1-3} 
\multirow{3}{*}{SPEC 2006 \cite{Henning2006}}
    & Mcf & 73 \\ \cline{2-3}
    & Cactus & 60 \\ \cline{2-3}
    & Astar & 9 \\ \hline
\multirow{3}{*}{PARSEC \cite{bienia2008parsec,bienia2009parsec}}
    & Freqmine (frqm) & 16 \\ \cline{2-3}
    & Canneal (canl) & 57 \\ \hline
\multirow{3}{*}{Intel GAP \cite{beamer2015gap}}
    & Betweenness Centrality (bc) & 113 \\ \cline{2-3}
    & Connected Components (cc, ccsv) & 56, 130 \\ \cline{2-3}
    & Single-Source Shortest Paths (sssp) & 144 \\ \hline
\multirow{1}{*}{Mantevo \cite{heroux2016mantevo}}
    & Path Finder (pf) & 41 \\ \hline
\multirow{3}{*}{NAS \cite{bailey2011parallel}}
    & DC & 49 \\ \cline{2-3}
    & MG & 99 \\ \cline{2-3}
    & SP & 141 \\ \hline
\end{tabular}
\vspace{-4mm}
\end{table}

\vspace{-3mm}
\section{Results}
\label{sec:results}
The goal of DeACT is to provide security from other tenants in decoupled FAMs without significantly impacting the performance. Hence we compare DeACT with two baselines, E-FAM and I-FAM. E-FAM is not secure but has better performance. I-FAM is secure but performs poorly (remember that I-FAM is similar to the optimization proposed by \cite{bhargava2008accelerating}). 

\subsection{FAM Address Translation Hit Rate}
\begin{figure}[hbt!]
    \centering
    \includegraphics[width=1\columnwidth]{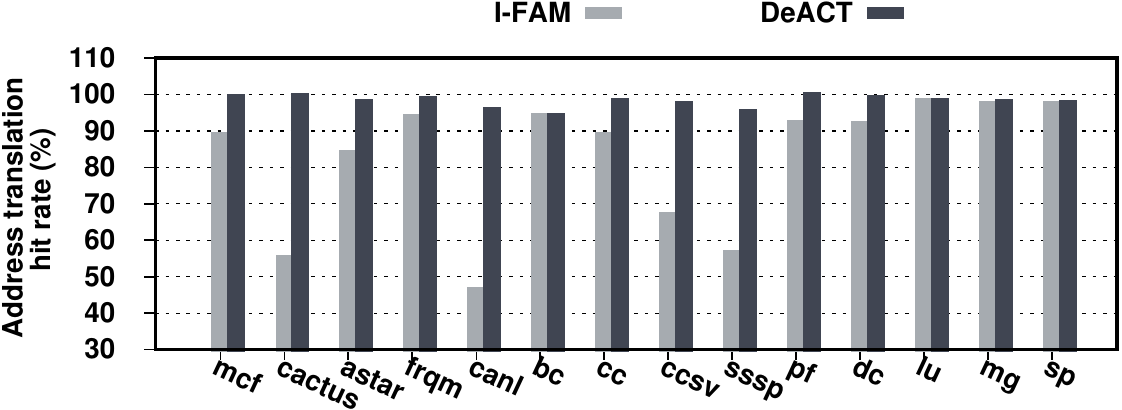}
    \caption{Address translation hit rate in I-FAM and DeACT}
    \label{fig:vm-add-tr-hits}
    \vspace{-2mm}
\end{figure}

Figure \ref{fig:vm-add-tr-hits} depicts address translation hit rate while accessing FAM in I-FAM and DeACT. The hit rate corresponds to the number of mapping entries that are cached in I-FAM and DeACT. DeACT has significantly higher hit rate (more than 90\%) because the FAM translation cache in local memory can cache significantly higher number of mapping entries than limited entries that can be cached in I-FAM using STU cache. For instance, hit rate for \textit{canl} benchmark is as low as 46.44\% in I-FAM. However, with DeACT the hit rate is improved to almost 95.88\%. Hence, only 4.12\% of the FAM accesses require page table to be walked.

\vspace{-4mm}
\vspace{-3mm}

\subsection{Address Translation Requests at FAM}
\begin{figure}[hbt!]
    \centering
    \includegraphics[width=1\columnwidth]{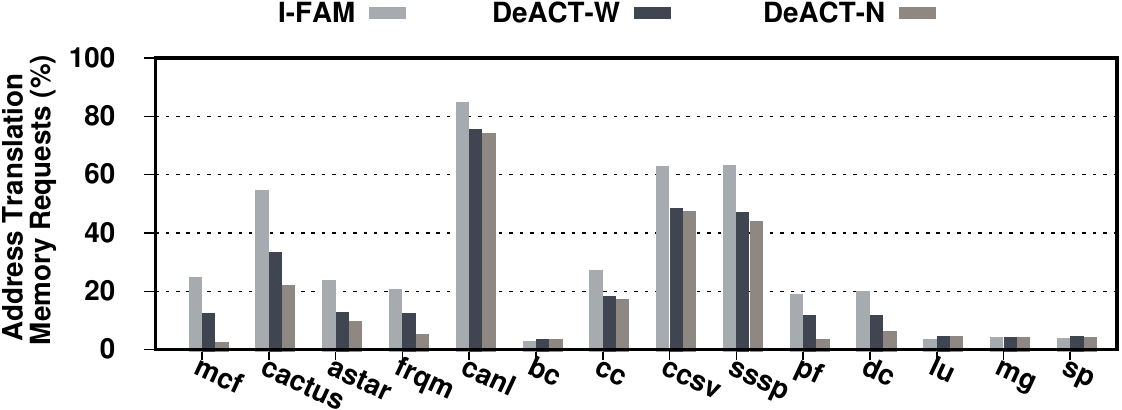}
    \caption{Percentage of address translation requests at FAM.}
    \label{fig:vm-sdma-mem}
    \vspace{-4mm}
\end{figure}

Although the frequency at which PTW is reduced with DeACT, it introduces an additional memory access for ACM. ACM is cached at STU cache and address translation is cached in the local memory. Hence in DeACT address translation and ACM has different hit rates. As shown in Figure \ref{fig:vm-rand-acm}, ACM hit rate in DeACT-W is not improved compared to I-FAM due to poor spacial locality. Hence the reduced number of address translation requests observed at FAM in DeACT-W, Figure \ref{fig:vm-sdma-mem}, is only due to reduced frequency of walking the page table and it also includes additional memory access for ACM. However, ACM hit rate is improved in DeACT-N due to non-contiguous caching of ACM. We observe address translation requests sent by the node to the FAM are reduced from 23.97\% to 11.82\% with DeACT-W, and this further reduces to 1.77\% with DeACT-N.
\vspace{-3mm}

\vspace{-3mm}
\subsection{Impact of DeACT on Performance}
\begin{figure}[hbt!]
    \centering
    \includegraphics[width=1\columnwidth]{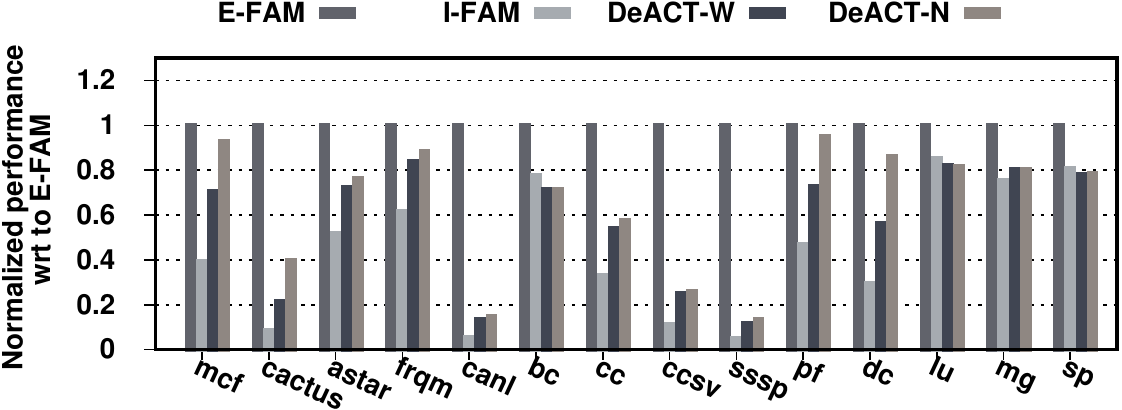}
    \caption{Normalized performance with respect to E-FAM.}
    \label{fig:sdma}
    \vspace{-4mm}
\end{figure}

In this section, we show how DeACT performs compared to E-FAM and I-FAM. E-FAM performs better than I-FAM and DeACT, hence we show our results with respect to E-FAM in Figure \ref{fig:sdma}. As mentioned in Section \ref{sec:motivation}, I-FAM slows down the system performance significantly. Our experiments demonstrate that DeACT can potentially bridge the gap between E-FAM and I-FAM. For instance \textit{mcf} slows down by 0.39x in I-FAM compared to E-FAM. DeACT-W performance is 0.7x wrt E-FAM, improving the performance by 1.79x compared to I-FAM. Further with DeACT-N the performance is improved by 2.55x and is just 0.92x times slower than E-FAM. This improvement is attributed to the increased FAM address translation hits, using local DRAM and increased ACM hits in STU, leading to decreased accesses to FAM for page table requests by the node, as shown in Figure \ref{fig:vm-sdma-mem}. The inequality between the performance improvement and reduction in observed percentage of address translation requests at FAM is because the local memory is accessed for every FAM access for the translation. 

\vspace{-2mm}

For \textit{canl, ccsv and sssp} benchmarks we observe a significant percentage of FAM address translation misses in I-FAM, and hence we observe increase in the percentage of address translation requests to FAM. The performance for such benchmarks even with DeACT-N is slower compared to E-FAM, 0.14x for \textit{canl}. However, compared to I-FAM, DeACT-N achieves a speed up by 2.7x for these benchmarks.

DeACT-N achieves a maximum performance improvement of 4.6x for \textit{cactus} benchmark. However, DeACT either does not improve or degrades the performance for \textit{bc}, \textit{lu}, \textit{mg} and \textit{sp} benchmarks. Because these benchmarks are very less sensitive to indirection in I-FAM, Figure \ref{fig:vm-sdma-mem}, as they have better address translation hit rate, Figure \ref{fig:vm-add-tr-hits}. However, in DeACT the DRAM has to be accessed for address translations, which is costlier than accessing STU cache. Also, the benchmarks have to go through two serial steps for address translation and access verification, unlike a single step for the same in I-FAM. Hence DeACT is better suitable for benchmarks which have a significant impact on performance with I-FAM. In total we observe an average performance drop of 69.7\% with I-FAM and with our proposed mechanism the performance degradation is 35.3\% compared to E-FAM. Hence, DeACT improves I-FAM by 80\%. 

\vspace{-1mm}
\subsection {Sensitivity Analysis}
The impact of performance in FAM systems is dependent on various factors. In this section, we show how DeACT behaves under various system configurations. The default system parameters are as shown in Table \ref{tab:simulationparameters}. Note that for sensitivity results we show geometric mean of the evaluated SPEC, PARSEC and GAP benchmarks separately.
Also, among NPB benchmarks, we observed \textit{dc} is the only benchmark which has significant performance impact in I-FAM even under various circumstances. Hence going forward we show sensitivity results only for \textit{dc} benchmark among NPB benchmarks. Also, since DeACT-N improves the performance more than DeACT-W, we focus on DeACT-N scheme. 

\subsubsection{STU Cache Size and Associativity}
\begin{figure}[hbt!]
    \centering
    \includegraphics[width=1\columnwidth]{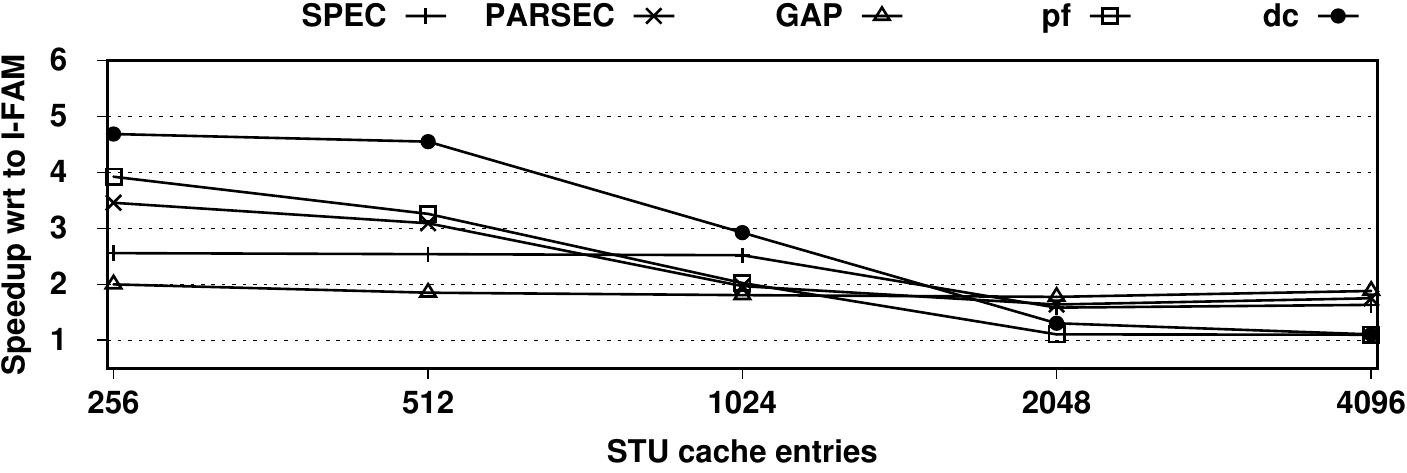}
    \caption{Performance improvement wrt STU cache size.}
    \label{fig:vm-sens-stu-size}
    \vspace{-4mm}
\end{figure}

\vspace{-0mm}
One of the main factors which impact the performance of I-FAM is the size of STU cache. STU is a hardware maintained outside the node to enforce system access control and page mapping. The number of entries STU can cache is limited, since we are proposing STU per node and is implemented in the routers connected to the nodes. Adding more entries indicates adding more hardware which increases the hardware budget and complicates routers. In our experiments STU caches 1024 entries. However, we study DeACT by varying STU cache size from 256 entries to 4096 entries. Figure \ref{fig:vm-sens-stu-size} shows performance speedup compared to I-FAM by varying STU cache size. As STU cache size decrease the speedup with DeACT is significantly high, 4.68x with 256 entries for \textit{dc} benchmark. However, as the cache size increase the performance improvement is confined which is obvious. The speedup reduces from 3.45x to 1.75x when STU cache size is varied from 256 entries to 4096 entries for PARSEC benchmarks. Higher STU cache size has higher hit rate and hence less address translation requests to FAM, however, higher cache size leads to more hardware overhead.

\vspace{-4mm}

Although we do not show here, we also evaluated DeACT by varying STU cache associativity. We observed that as the associativity increases, the performance improvement with DeACT decreases and gets saturated. When associativity is four the performance improvement is 3.26x for \textit{dc} benchmark and is 2.66x when associativity is 32. The speedup is 2.5x when associativity is greater than 32 for the same benchmark. Similarly for PARSEC benchmarks the speedup is 2.18x, 1.83x and 1.81x when associativity is 4, 32 and greater than 32.


\subsubsection{Access Control Metadata Size}
\begin{figure}[hbt!]
    \centering
    \includegraphics[width=1\columnwidth]{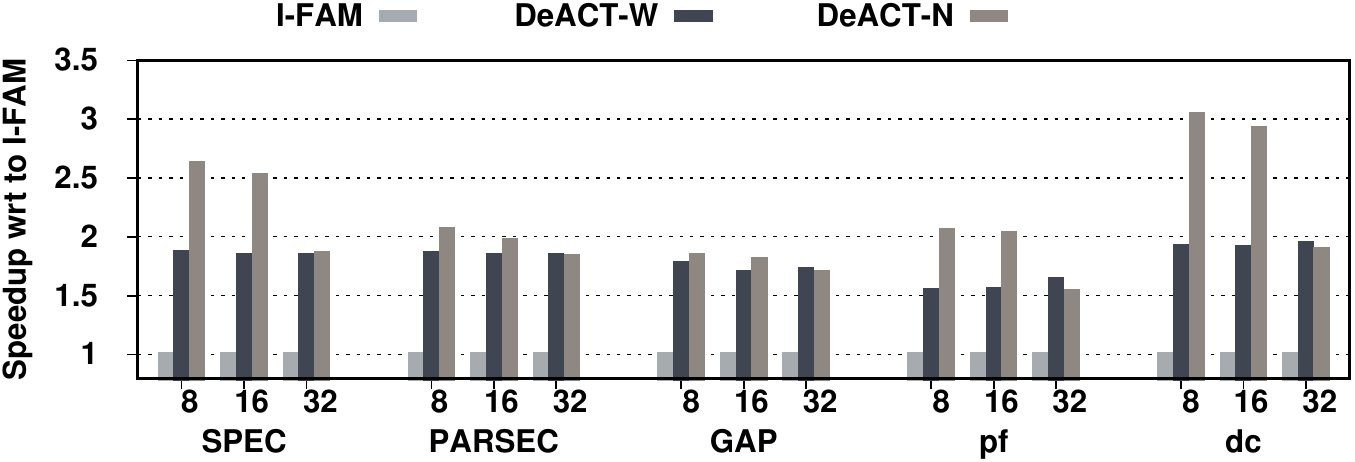}
    \caption{Metadata size effect on performance.}
    \label{fig:vm-sens-acbits}
    \vspace{-4mm}
\end{figure}

ACM size is a key design aspect of DeACT as the number nodes supported by FAM systems is dependent on metadata size, refer to Section \ref{sec:vm-deact-overview}. 
With 16-bit metadata size FAM systems hosts a total of 16383 nodes and with 8-bit metadata 8191 nodes are supported. 
When ACM is 8 bits STU in DeACT-W can cache metadata of eight consecutive pages, with 16-bit metadata STU can cache metadata of four consecutive pages, and with 32-bit metadata STU can cache metadata of two consecutive pages, increasing the amount of metadata cached by 8x, 4x and 2x respectively compared to I-FAM. However, we observe that the performance improvement is almost same for these three scenarios, Figure \ref{fig:vm-sens-acbits}. This is because, as asserted, although caching of ACM increases in DeACT-W, it caches only ACM of contiguous pages and since allocation of FAM is random excess caching of ACM is not leveraged.

\vspace{-0mm}

When ACM size is 8 bits, tag and metadata pair in DeACT-N requires 52 bits (44 bits for tag and 8 bits for ACM, see Section \ref{ssec:vm-deact-n}). Hence in a single way, STU cache can store two tag and ACM pairs, similar to when ACM is 16 bits. However, the amount of memory required to store ACM of all the pages is reduced to half. As an experimental model we further reduce the size of the tag to allocate three pairs of tag and ACM per STU cache way, when ACM is 8 bits. When ACM is 32 bits STU can cache only one tag and ACM pair in DeACT-N. We observe that as caching of tag and ACM pairs in STU cache way increase, from one to three, the performance improvement with DeACT-N also improves. For instance. the system performance improves by 2.62x, 2.52x and 1.85x when one, two and three pairs of tag and ACM are cached in each STU cache way, for SPEC benchmark. It is interesting to note that when only one pair of tag and ACM is stored in STU cache way the performance improvement is less than or equal to DeACT-W. This is because when only one pair of tag and ACM is cached in each STU cache way the performance improvement is only due to increased address translation hits in FAM translation cache and ACM hit rate is same as I-FAM in DeACT-N.

\subsubsection{Fabric Latency}
\vspace{-0mm}
\begin{figure}[hbt]
    \centering
    \includegraphics[width=1\columnwidth]{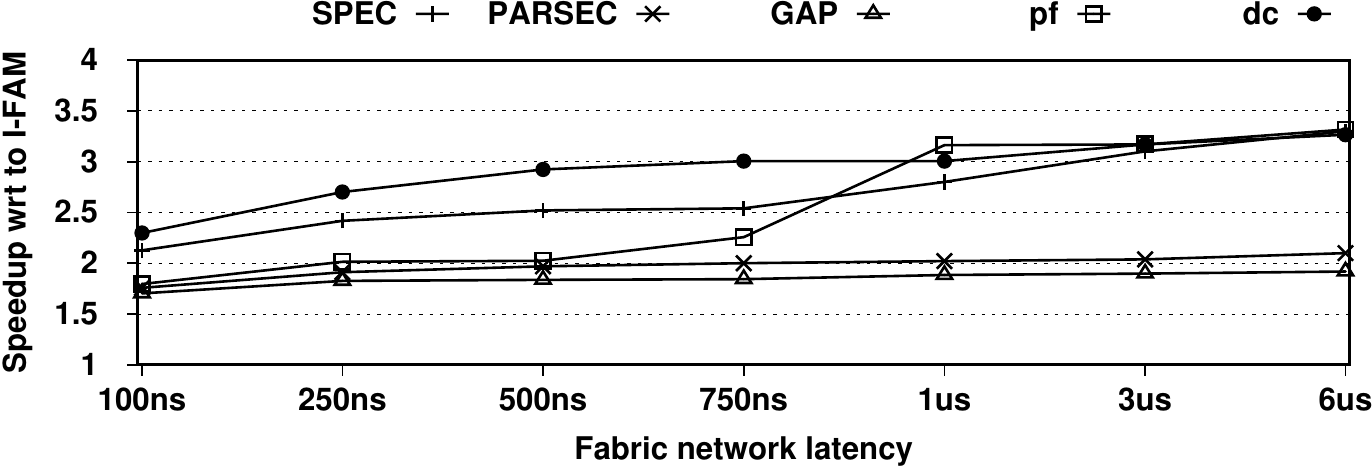}
    \caption{The impact of fabric latency on performance.}
    \label{fig:vm-sens-fabric-latency}
    \vspace{-6mm}
\end{figure}
We propose DeACT for FAM architectures specifically and hence one of the crucial parameter to consider for such architectures is fabric network latency. In our approach we considered 500ns as fabric latency. However, fabric networks are being explored intensively by various fabric providers \cite{gen2019gen,sharma2019compute,ccix,opencapi2018opencapi}. Previous approaches considered various fabric network latencies \cite{gao2016network,shan2018legoos,aguilera2017remote,kommareddy2019page}. Thus we evaluated DeACT under the influence of various fabric latencies, Figure \ref{fig:vm-sens-fabric-latency}. An obvious observation is that when fabric network latency is less the performance improvement with DeACT is also less and when the network latency is high the performance improvement with DeACT is more. This is because when the fabric network latency is less, performance degradation of I-FAM itself is less, compared to E-FAM. This goes inversely when fabric latency is high. We see that even when fabric network latency is less, 100ns, DeACT achieves an improvement of 1.79x wrt to I-FAM. In contrast, when the network latency is 6us DeACT speeds up I-FAM by 3.3x for \textit{pf} benchmark.

\subsubsection {Number of Nodes}
\label{ssec:nodes}
The interconnecting fabric connects multiple PEs to the decoupled FAM modules. A single FAM module is expected to be part of a single memory pool. FAM architectures are constructed with multiple such memory pools and PEs. The performance of FAM architectures depends on the number of PEs and number of memory pools. We maintained memory pools directly proportional to the number of nodes and each node has four PEs. For instance, an eight node system consists of 32 PEs and eight FAM modules. Each of the PEs accesses any of the memory pools. Memory pools and the PEs are connected though a common fabric network. The delay in accessing FAM depends on the number of nodes sharing the fabric interface and memory. Although scalability is beyond the scope of this paper we evaluated our approach when multiple nodes (up to 8) share the fabric. As the number of PEs sharing the fabric increase, we observe that the slowdown due to I-FAM is higher. This is due to more cycles are consumed to fetch FAM page table entries since fabric network and memory are shared by the nodes. As a result the performance improvement with DeACT is more since DeACT avoids accessing FAM for page table entries, for most of the time. When the fabric network and memory is allocated to only one node the performance improvement with DeACT is 2.92x for \textit{dc} benchmark and it increases to 3.26x when fabric network and memory are shared between 8 nodes.

\vspace{-3mm}
\begin{figure}[hbt]
    \centering
    \includegraphics[width=1\columnwidth]{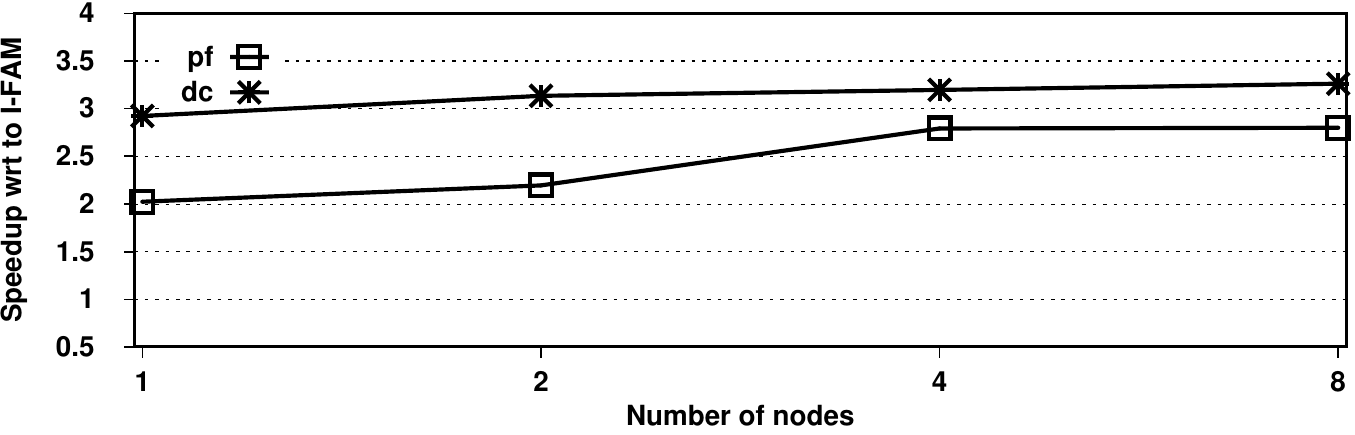}
    \caption{Impact of increasing the number of nodes on performance.}
    \label{fig:vm-sens-stu-alloc}
    \vspace{-6mm}
\end{figure}

\section{Discussion}
\label{sec:disc}
\vspace{-2mm}
We proposed large pages (1GB) in the global memory for only shared pages to reduce the memory occupied by the shared pages access control bitmap. Ideally large pages would also reduce address translation overhead. But for FAM systems this is not completely true. Hence we shed some light to discuss how large pages for shared memory is achievable and would not provide benefits for non-shared global memory. Also, we discuss how to invalidate system-level translations and ACM to migrate jobs smoothly between the nodes by introducing logical node numbers for the jobs in FAM systems. We even discuss the scalability and emerging fabric network aspects for our approach in FAM systems.


\noindent \textbf{Shared Pages:} FAM systems allow for easy sharing of data by registering to a common memory region in the global memory, shared pages. The proposed access control bitmap mechanism for shared pages is also applicable for baseline I-FAM wherein the address translations are not decoupled from the access control. Since for baseline I-FAM as well the system-level access control alone is not sufficient to indicate how many pages the page is shared with. System-level access control can only indicate if the page is shared or not. Also, considering the fact that the modern applications share a lot of data frequently, large shared pages, as proposed, would lessen the burden of memory manager to keep a track of the shared pages.

\noindent \textbf{Page Migration:} A requirement for hybrid cloud systems is to migrate jobs between the nodes\cite{foster2008cloud}. With I-FAM it is easy to move jobs to another system, since host/system data need not be moved. Only overhead is to invalidate node addresses of the job in both node (value) and fabric (tag) page tables. With DeACT method, node-level translations are invalidated similar to I-FAM. However, system-level translations are invalidated by a) accessing fabric translation cache at DRAM b) invalidating access control at STU cache and c) updating access control of migrating job with the target node at the global memory. Hence, the overhead of system-level mapping shootdown is accessing global memory separately to update access control of migrating jobs and excess DRAM writes to invalidate system-level mappings at in-memory fabric translation cache. 

\noindent To allow smooth migration of jobs, logical node IDs per jobs can be used. Logical node IDs are assigned to a jobs by the resource managers \cite{computing2015torque,greenbergredfish}. For such a scheme migrating jobs requires only assigning new logical node ID in the destination node and removing assigned logical node ID in the old node, apart from invalidating page mappings.

\noindent \textbf{Scalability:} As asserted although scalability is beyond the scope of this paper we show performance improvements for 8 nodes with DeACT approach. We show that as the number of nodes sharing Gen-Z fabric increase the improvement with DeACT also increases since most of the system-level address translations avoid fabric network traffic and contention at FAM generated by other nodes in our approach. As such our hypothesis is valid even when more number of nodes share fabric interconnect.


\noindent \textbf{Large Pages and Memory Regions:} While using large pages improves system performance, since TLB caches page mapping for larger memory, it has some drawbacks \cite{gaud2014large,talluri1992tradeoffs}. FAM systems has separate concerns in addition to the concerns with large pages in traditional in-node memory systems. To improve node proximity of data, pages are placed near the computing units, in local memory. With large pages there are three crucial concerns to achieve node proximity of data a) As local memory size is limited it can host less number of large pages which covers lesser number of applications. b) Applications do not frequently use the entire large page, but access only specific portions of the large page frequently. This results in poor utilization of the DRAM which is critical for disaggregated memory systems. c) Identifying frequently accessed pages is tricky as large pages cover a number of small pages and frequently used small pages can get scattered between multiple large pages. Shared pages cannot have node proximity as these pages have to be placed in the FAM to get accessed by various nodes. Hence large pages for shared pages would not generate additional concerns unlike pages which are not shared.




\vspace{-0mm}
\section{Related Work}
\label{sec:related}
\vspace{-2mm}
Recently, disaggregating memory from PEs has been explored as an alternative memory architecture to overcome various operational and scalability challenges of in-node memory architectures \cite{shan2018legoos,aguilera2017remote,aguilera2018remote,lim2009disaggregated, lim2012system,kommareddy2019page}. Works such as \cite{gu2017efficient,gen2019gen,ccix,birrittella2015intel} discuss and explore fast interconnect to enable decoupling memory. However, there has been limited work discussing virtual memory and security for such systems. Lim et al. \cite{lim2009disaggregated} discussed two stage address translations for FAM systems, but their approach is limited to using remote memory merely as a swap space. Lim et al. \cite{lim2009disaggregated} also proposed fine-grained remote memory accesses, which is similar to E-FAM and is not secure as discussed. Shan et al. \cite{shan2018legoos} proposed decoupled OS for FAM systems and the address translations are performed by the FAM modules. However, for such a scheme to work the caches have to be virtually indexed and virtually tagged which is not adopted and is not a practical design. Also it requires significant amount of changes to the OS. Aguilera et al. \cite{aguilera2017remote} invalidated virtual memory paging for such huge memory designs.
Aguilera et al. also proposed fixed virtual address regions for the nodes \cite{aguilera2018remote}. However, this requires modifications to applications' binaries. In this paper, we discuss virtual memory support for FAM architectures with two stage address translations (I-FAM) and propose DeACT scheme to accelerate address translations. 

Decoupling access control and address mappings have been explored previously. Alam et al. \cite{alam2017yourself} discussed decoupled address control from address mapping allowing applications to perform address translations by itself. Olson et al. \cite{olson2015border} proposed an approach to sandbox accelerators through providing them with flat address space. DeACT uniquely leverages the architecture layout of FAM architectures when decoupling access control from translation; it allows fast caching of translations in local nodes' main memories, and maintains access control in the trusted area (i.e., at system-level). Additionally, DeACT supports data sharing across nodes and leverages system-level translation units at the fabric. 



Although virtual machine guests are different from nodes in FAM systems, in both the cases virtual addresses are translated at 2 stages to access memory. A significant amount of work has been done to improve the performance of virtualized conventional machines. Bhargava et al. \cite{bhargava2008accelerating} accelerate 2D PTW by studying reuse of page entry references and extend PTW caches to temporarily cache nested dimension. Ahn et al. \cite{ahn2012revisiting} revisited hardware assisted page walks by speculative shadow paging mechanism, called speculative inverted shadow paging, which is backed by non-speculative flat nested page tables. The speculative mechanism provides a direct translation with a single memory reference for common cases and eliminates the page table synchronization overheads. Agile paging is proposed by Gandhi et al. \cite{gandhi2016agile}. Agile paging allows virtualized page walk to start with the shadow paging for stable upper levels of the page table and allows switching in the same page walk to nested paging for lower levels of page table which receive frequent updates. This way agile paging makes use of both shadow paging and nested paging. While these approaches improve the system performance, these are proposed for virtual machines and our approach is orthogonal to these schemes.

\vspace{-0mm}
\section{Conclusion}
\label{sec:conclusion}
\vspace{-2mm}
Disaggregated memory systems is a promising future architecture which provides a path for future memory systems by solving current memory concerns in scalability, with the sharing of huge data sets such as social graphs, as well as complex scientific data-sets in HPC. The challenges associated with their design in an HPC context is how to best utilize the resources of the system and balance these against the ever increasing demands to achieve better performance. To this end, we discuss virtual memory management in disaggregated memory systems and propose solutions to speed up address translations and provide security for such systems. While approaches like \cite{ahn2012revisiting,gandhi2016agile} reduce the number of memory accesses to fetch address mapping in virtualized systems, they target native virtualized systems. Due to the hierarchical nature of the memory, disaggregated memory systems have its own challenges supporting virtual memory.
We show that exposing global memory to the nodes needs OS alterations and compromise security from neighbour nodes. Virtual memory approach for disaggregated memory (indirect memory access) does not ask for OS modifications and provides security from neighbour nodes, but performs poorly. Although virtual memory support is discussed for disaggregated memory systems in \cite{lim2009disaggregated,shan2018legoos,aguilera2018remote,aguilera2017remote}, in such approaches remote memory is merely used as a swap space and required application and OS changes. We proposed decoupled address translation and access control metadata approach to improve the performance of I-FAM. We show that an improved spacial locality of system-level translations by decoupling the system-level address translations from system-level access control metadata and caching the decoupled FAM translations in the local memory. We also explore access control metadata caching in STU cache to improve the performance.
Overall, we achieved a performance improvement up to 4.59x (1.8\% on average).

\section{Acknowledgments}
This work has been funded through Sandia National Labora-tories (Contract Number 1844457) Sandia National Laboratories is a multi-mission laboratory managed and operated by National Technology and Engineering Solutions of Sandia, LLC., a wholly owned subsidiary of Honeywell International,Inc., for the U.S. Department of Energy's National Nuclear Security Administration under contract DE-NA0003525.

\bibliographystyle{IEEEtranS}
\bibliography{main}

\end{document}